\documentclass[12pt]{article}
\usepackage{latexsym,amsmath,amssymb,epsfig}

\newcommand{\bea}{\begin{eqnarray}}
\newcommand{\eea}{\end{eqnarray}}
\newcommand{\be}{\begin{equation}}
\newcommand{\ee}{\end{equation}}
\newcommand{\beast}{\begin{eqnarray*}}
\newcommand{\eeast}{\end{eqnarray*}}
\newcommand{\pkt}{\; .}
\newcommand{\kma}{\; ,}
\newcommand{\nn}{\nonumber}
\newcommand{\eqn}[1]{(\ref{#1})}
\newcommand{\cald}{{\cal D}}
\newcommand{\calg}{{\cal G}}
\newcommand{\calm}{{\cal M}}
\newcommand{\bfm}{{\bf M}}
\newcommand{\labelcaption}[2]{\caption[#1]{\label{#1}#2}}
\newcommand{\tir}{{\tilde r}}
\newcommand{\tr}{{\rm tr}}
\def\e{{\rm e}}
\begin{document}
\begin{titlepage}
\begin{flushright}
DO-TH-03/08 \\
hep-th/0307202 \\
July 2003
\end{flushright}

\vspace{20mm}
\begin{center}
{\Large \bf
One-loop corrections to the metastable vacuum decay}
\vspace{10mm}

{\large
J\"urgen Baacke $^\#$ \footnote{e-mail: baacke@physik.uni-dortmund.de}
and
George Lavrelashvili $^*$ \footnote{e-mail: lavrela@rmi.acnet.ge}}

\vspace{15mm}

{\large $^\#$ Institut f\"ur Physik, Universit\"at Dortmund \\
D - 44221 Dortmund, Germany\\
and \\
$^*$ Department of Theoretical Physics\\
A.Razmadze Mathematical Institute\\
GE - 0193 Tbilisi, Georgia}

\vspace{15mm}

\bf{Abstract}
\end{center}
We evaluate the one-loop prefactor in the false vacuum decay rate in a
theory of a self interacting scalar field in $3+1$ dimensions.
We use a numerical method, established some time ago,
which is based on a well-known theorem on functional determinants.
The proper handling of zero modes and of renormalization is discussed.
The numerical results in particular show that
quantum corrections become smaller away from the thin-wall case.
In the thin-wall limit the numerical results are found to join into those
obtained by a gradient expansion.
\end{titlepage}

\setcounter{page}{2}
\section{Introduction}
\par

First-order phase transitions play an important role
in various phenomena from solid state physics to cosmology.
The basic theoretical concepts of these transitions have been developed
long ago \cite{La67,La69,vko75,Co77,CaCo77,Co79}.
The phase transition proceeds via formation of stable phase
(or true vacuum) bubbles within a metastable (or false vacuum) environment,
 and subsequent growth of these bubbles.
Two mechanisms of the first order phase transitions are known:
quantum tunnelling and thermal activation.
In both cases the decay rate of a metastable state is given by the formula
\be                                                 \label{rate0}
\gamma = {\cal A} \e^{-{\cal B}} \pkt
\ee
For tunnelling in a $(3+1)$ dimensional theory the quantity $\cal B$
in the exponent
is given by the {\it classical} 4d Euclidean action evaluated on a bounce,
a finite action Euclidean solution of classical equations of motion,
asymptotically approaching the false vacuum.
For thermal activation at nonzero temperature $\cal T$ the
exponent is given by  $- \cal B=E/T$, where $\cal E$ is the energy of
a critical bubble (sphaleron), which is a static solution "sitting" on
a top of a barrier separating two vacua. The bounce as well as the sphaleron
are unstable solutions with just one negative mode. Bubbles smaller than
critical collapse, and the ones bigger than critical expand and lead to
the transition to a new phase. These static solutions and Euclidean
solutions are related, namely the sphaleron
in  $(d+1)$ dimensions can be viewed
as a bounce in $d$ dimensions.

The leading order estimate for the transition rate is easy to obtain,
it just requires solving - in general numerically - an ordinary, though
nonlinear differential equation. Analytic estimates can be obtained
in the so-called thin-wall approximation.

The pre-exponential factor ${\cal A}$ in  Eq.~(\ref{rate0})
is calculated taking into account
quadratic fluctuations about the classical solution and is
given as a ratio of the functional determinants. In general
it is a very difficult task to calculate analytically the determinants,
while the background solution itself is not known in a closed form.
It has taken two decades until the first (numerical) computations
of the quantum corrections to the leading order semiclassical transition
rates have appeared \cite{ks84,ks86,se88,baki93,Carson:1990jm,baju93}.
Of course nowadays the CPU
time requirements for such computations are, even for more
involved systems, of the order of seconds. On the other hand
the requirements of a precise renormalization,
which compares exactly to the one of perturbative
quantum field theory, and of the inclusion and careful
treatment of high partial waves, have of course
remained the same. The method used here has been developed and tested
for various systems and has become a standard procedure.
It is well suited for computations of coupled channel
problems as well \cite{Baacke:1995bw}.

While the special technique used here applies only to
the computation of functional
determinants, the general approach  can be used as well for
computing zero point energies \cite{ba91,ba92} via Euclidean
Green functions. Of course functional determinants can
be computed likewise using  Euclidean Green functions
\cite{baju93,Baacke:1994bk}. Various other techniques for computing
the exact quantum corrections have been developed in the past decade.
In Refs.  \cite{Carson:1990jm,Diakonov:1995xz} the heat kernel is computed using
a discretisation of spectra, in Ref. \cite{Graham:2002xq}  Minkowskian
instead of Euclidean Green functions are used, and in Ref. \cite{Bordag:2002sa}
the zero point energy is computed via the $\zeta$ function.

The effective action may be computed approximatively by using
gradient expansions. There is an ample literature on this subject.
We just quote Refs. \cite{Hellmund:1993uh,Fliegner:1993wh,Fliegner:1997rk,mr99}
for expansions using advanced heat kernel techniques, and Ref.
\cite{Caro:1993fs} for expansions based on Feynman graphs.

The leading quantum corrections, being essentially a one loop effect,
can be viewed as a ``summary" of the particle creation during
the phase transition \cite{rubakov84}.
The question about the quantum corrections is very important one,
while there are cases when particle creation is so strong that
it drastically modifies the original classical tunnelling solution
\cite{surig97,levkov02}.

The aim of the present paper is to calculate the pre-factor $\cal A$
for tunnelling transitions in a theory of one self-interacting
scalar field theory in (3+1) dimensions.

The rest of this paper is organized as follows:
In the next Section we will describe our strategy for calculation
of one loop effective action.
In Section \ref{treelevel} we formulate our model, specify the form of
the potential, write the equation of motion for the bounce
and present our numerical results for classical action $S[\varphi]$.
In Section \ref{determinant} we describe the calculation of the
fluctuation determinant, Eq. (\ref{det}). There we also discuss
regularization and renormalization.
Our numerical results are presented and discussed in Section \ref{num}.
We end with some general remarks and conclusions in Section \ref{concl}.
Formulas describing the thin-wall approximation and gradient expansion
are collected in the Appendixes A and B respectively.

\setcounter{equation}{0}
\section{General strategy} \label{general}
\par

We will consider phase transitions in a theory of
one self interacting scalar field $\varphi$ in $3+1$ dimensions.
Corresponding Euclidean action is
\be                                                 \label{S}
S[\varphi]= \int d^4 x \left( \frac{1}{2}
\left( \partial _\mu \varphi \right) ^2 + U(\varphi ) \right)\kma
\ee
where the field potential $U(\varphi )$ is assumed to have two
non-degenerate minima $\varphi=\varphi_{-}$ and
$\varphi=\varphi_{+} >0$ (compare Fig. \ref{potU}) and
it will be given explicitly in the next section.
For convenience we have fixed the value of $\varphi$ in the
{\em unstable} vacuum as $\varphi_-=0$.

Any state built on the local minimum $\varphi_-$ is metastable.
It can tunnel locally towards the $\varphi_+$ phase.
The tunnelling rate per unit volume per unit time, $\gamma =\Gamma /VT$,
is supposed to be dominated by the classical action $S_{cl}$ of a
field configuration, the bounce $\varphi_{b}(x)$,
which looks like a bubble of the $\varphi_+$-phase within the
$\varphi_-$ phase.  In particular it can be shown \cite{CGM78}
that the bounce configuration $\varphi_{b}(x)$ which minimizes the action
is spherically symmetric in four-dimensional Euclidean space.
In the tree level approximation the decay rate is determined essentially by
the tunnelling coefficient,
$\gamma \propto \exp \{ -S_{cl}[\varphi_{b} (x)] \}$\footnote{For a more
concise statement see Section \ref{num}.}.

The tree level tunnelling rate receives corrections in higher orders
of the semiclassical approximation. In quantum field theory
the fluctuations around the bounce contribute in the
next-to-leading order approximation a pre-exponential factor to the decay rate.
The rate per volume and time is known to take the form
\cite{CaCo77}
\be                                               \label{rate}
\gamma = \left( \frac{S_{cl}[\varphi]}{2\pi}\right)^2
\left| {\cal D} \right| ^{-1/2} \exp \left\{-S_{cl}[\varphi]
-S_{ct}[\varphi]\right\}
\ee
to one-loop accuracy. The coefficient ${\cal D}$ here is defined as
\be                                                \label{det}
\cald[\varphi]
\equiv\frac{\det'( -(\partial /\partial \tau )^2 -\Delta + U''(\varphi))  }
{\det(-(\partial /\partial \tau )^2 -\Delta + U''(0))  }
=\frac{\det'(\calm)  }{\det (\calm^{(0)})} \pkt
\ee
The prime in the determinant implies omitting of the four
translation zero modes. With the second equation we have introduced
the fluctuation operator in the background of the bounce
\be
\calm =-(\partial /\partial \tau )^2 -\Delta + U''(\varphi)
\ee
and its counterpart $\calm^{(0)}$ in the unstable vacuum.

The counterterm action $S_{ct}$ is necessary in order to
absorb the divergences of the one-loop effective action
\be
S^{eff}_{1-loop}[\varphi]= \frac{1}{2}\ln | \cald[\varphi] | .
\ee

In order to evaluate the one loop effective action
we decompose fluctuations about the bounce $\varphi_b$ into
$O(4)$ spherical harmonics,  calculate the ratio of determinants $J_l$
of partial wave fluctuation operators  and
obtain $\ln {\cal D}$ as $\sum_l d_l\ln J_l$, where
$d_l$ is the $O(4)$ degeneracy $d_l=(l+1)^2$ (see e.g. \cite{cw84}).
In calculating $\ln {\cal D}$ we exclude the divergent perturbative
contributions of first and second order in the external field
of the bounce $\varphi_b$. The regularized values of
these contributions are then added analytically.
All divergences of $\ln {\cal D}$ appear in the standard tadpole and
fish diagrams.
We will not specify $S_{ct}$ explicitly, we will equivalently omit
the divergent parts of $\ln \cald[\varphi]$  using the $\overline{MS}$
convention.


\setcounter{equation}{0}
\section{The Tree-Level Action} \label{treelevel}
\par

In this section we specify our model, discuss the bounce solution
and properties of corresponding classical action.
We parameterize the $\varphi^4$-potential with two minima as
\be                                                \label{U}
U(\varphi)=\frac{1  } {2  } m^2 \varphi^2 - \eta \varphi^3
+ \frac{ 1 } { 8 } \lambda \varphi^4 \kma
\ee
and choose the same dimensionless variables as in
Ref. \cite{DLHLL,baki93}:
$x^\mu =X^\mu /m$ for $\mu=0,1,2,3$, and $\varphi = \frac{m^2}{2\eta} \Phi$.
The classical action then takes the form
\be                                                  \label{scl}
S_{cl}(\varphi)= \beta \tilde{S}_{cl}(\varphi) \kma
\ee
where rescaled classical action $\tilde{S}_{cl}(\varphi)$ is
\be                                                  \label{E}
\tilde{S}_{cl}(\varphi)= \int d^4 X \left(\frac{1}{2}
\left( \nabla \Phi \right) ^2 + U(\Phi) \right) \; ,
\ee
with
\be                                             \label{pot}
U(\Phi)= \frac{1}{2} \Phi^2 - \frac{1}{2} \Phi^3
+ \frac{\alpha}{8} \Phi^4 \kma
\ee
and $\alpha$ and $\beta$
\be                                              \label{alpha}
\beta=\frac{m^2}{4\eta^2},~\qquad   \alpha = \lambda \beta .
\ee
being two dimensionless  parameters
\footnote{We use $\hbar=c=1$ units throughout this paper.}.
Parameter $\alpha$ varies from 0 to 1 and
controls the strength of self--interaction and
shape of the potential. For $\alpha=0$ the second minimum disappears,
whereas in the limit $\alpha \to 1$ two minima become degenerate
(see Fig. \ref{potU}). Parameter $\beta$ controls size of the loop corrections.
In order semiclassical approximation to be valid $\beta$ should not be
too small (see Section \ref{num} for details).

The bounce is non-trivial, $O(4)-$ symmetrical stationary point of
$S_{cl}$, Eq.~(\ref{E}), obeying the Euler -- Lagrange equation
\be                                                \label{eq}
\frac{d^2\Phi  } {dR^2  }+\frac{ 3 } { R } \frac{d\Phi  } {dR  }
-\Phi +\frac{3}{2} \Phi^2 -\frac{\alpha}{2} \Phi^3 =0 \; ,
\ee
and boundary conditions
\be
\frac{d\Phi}{dR}|_{R=0}=0,\qquad \Phi_{R\to\infty}=\Phi_{-} \pkt
\ee
Here $R=((X^0)^2+|\vec{X}|^2)^{1/2}$.
The equation (\ref{eq}) at least for not very big $\alpha$
can be easily solved  numerically, e.g., by the shooting method.
We display some profiles $\Phi (R)$ in Fig. \ref{phi}
for various values of the parameter $\alpha$ .

The classical action $\tilde{S}_{cl}(\varphi)$ as a function
of $\alpha$ is plotted in Fig. \ref{sclass} (left).
For small $\alpha$ classical action $S$ goes to a constant and
$\tilde{S}_{cl}(\alpha=0)= 90.857 $.
In the limit $\alpha \to 1$ the thin-wall case is realized
(see Appendix A) and the classical action diverges as
$(1-\alpha)^{-3}$. The ratio of the classical
action computed numerically to the analytic thin-wall expression
\be
 \tilde{S}^{tw}_{cl} = \frac{\pi^2}{3 (1-\alpha)^3}
\ee
is displayed in Fig. \ref{sclass} (right). It
tends to unity for $\alpha \to 1$, as it should.
Note, that the  radius of the bounce increases rapidly in this limit and
numerical calculations become delicate. So, in the present article
we restrict ourselves to the interval $\alpha\in [0,0.95]$.


\setcounter{equation}{0}
\section{Calculation of the Fluctuation Determinant}    \label{determinant}
\par

In this section we discuss a method of computing the ratio of
functional determinants (\ref{det}) which is based on earlier papers
\cite{ks84,se88,baki93}.

The explicit form of the operator in the nominator (\ref{det}) is
\be                                                  \label{H}
{\cal M} = -\Delta_4+m^2 + V(r) \pkt
\ee
Here $\Delta_4$ is the 4-dimensional Laplace operator, and
we have introduced the potential $V$ as
\bea                                                \label{V}
V(r) = U''(\varphi )-m^2
= -6\eta\varphi (r)+\frac{3}{2}\lambda \varphi^2(r) \nn \\
= m^2 \left[- 3  \Phi(R) +\frac{3}{2}\alpha \Phi^2(R)\right]
\equiv m^2 V(R) \pkt
\eea
The ``free'' operator $\calm^{(0)}$, corresponding to the metastable phase
where $\varphi=0$ and where $ m^2=U''(\varphi=0)$ takes the
same form as (\ref{H}), but with $V(r)=0$.

Due to the  $O(4)$ spherical symmetry of the bounce
the operators $\calm$ and $\calm^{(0)}$ can be separated
with respect to $O(4)$ angular momentum. We introduce the
partial wave operators
\be                                                  \label{Hl}
\bfm_l (\nu)=-\frac{d^2  } {dr^2  } - \frac{ 3 } {r  } \frac{d  } {dr  }
+ \frac{l(l+2)}{r^2}+ \nu ^2 + m^2 + V(r)
\kma
\ee
with an additional variable $\nu$ that will be used later on.
In terms of these operators we can write
\be                                              \label{prod}
{\cal D}[\varphi]\equiv
{\prod_{l,n}}'
\left[\frac{\omega^2_{ln}}{{\omega^{{}2}_{l n {(0)} }}} \right]
=\prod_{l=0}^\infty \left[\frac{\det' \bfm_l(0) }
{\det \bfm_l^{(0)}(0)} \right]^{d_l} \kma
\ee
where $d_l$ is the degeneracy of the $O(4)$ angular momentum,
$d_l=(l+1)^2$. Prime denotes that for $l=1$ we have to remove the four
translational zero modes.

The ratio of determinants of the radial operators
\be                                                 \label{Dl}
J_l(\nu) = \frac{\det \bfm_l(\nu) }{\det \bfm_l^{(\nu)}(0)}
={\prod_{n}}
\left[\frac{\omega^2_{ln}+\nu^2}{{\omega^{{}2}_{l n {(0)} }+\nu^2}} \right]
\ee
can be computed using the theorem on functional determinants as
described in the next section. Note that $\omega^2_{ln}$ always denotes
the eigenvalues of $\bfm_l(0)$, or more generally the
eigenvalues of $\calm$, the analogous definition holds
for $\omega^{{}2}_{l n {(0)} }$.


\subsection{Determinants of the Radial Operators}                \label{s1}
\par

In order to find $J_l(\nu )$ (\ref{Dl}) we make use of a known theorem
\cite{theorem,Co79} whose statement is
\be                                                 \label{th}
\frac{\det \bfm_l(\nu)}{\det \bfm_l^{(0)}(\nu) } =
\lim_{r\to\infty} \frac{\psi_l(\nu,r)}{\psi_l^{(0)}(\nu,r)} \pkt
\ee
Here $\psi_l(\nu,r)$ and $\psi_l^{(0)}(\nu,r)$ are solutions to equations
\be           \label{eth}
\bfm_l(\nu)\psi_{\nu,l}=0 \pkt \qquad
\bfm_l^{(0)}(\nu)\psi_{\nu,l}^{(0)}=0 \kma
\ee
and have same regular behavior at $r=0$. More exactly,
the boundary conditions at $r=0$ must be chosen in such a way that
the right-hand side of Eq. (\ref{th}) tends to 1 at $\nu\rightarrow\infty$.

It is convenient to factorize the radial mode functions into the solution
$\psi_l^{(0)}(\nu,r)$ for  $V(r)=0$ and a factor $1+h_l(\nu,r)$
which takes into account the modification introduced by the potential.
If $V(r)$ is of finite
range the functions $\psi_l^{(0)}(\nu,r)$ and $\psi_l(\nu,r)$ have the same
behavior near $r=0$ and as $r \to \infty$. Near $r=0$ they behave as
$r^l$ and as $r\to \infty$ they behave as $\exp(-\kappa r)$ where
$\kappa=\sqrt{\nu^2+m^2}$. Furthermore the requirement of analogous behavior
near $r=0$ introduces the initial conditions $h(0)=h'(0)=0$.
The function $h(r)$ then simply starts from zero at $r=0$ and goes
smoothly to a finite constant value $h_l(\nu,\infty)$ as
$r\to \infty$.   The solutions $\psi_l^{(0)}(\nu,r)$ are given
in terms of modified Bessel functions as
\be
\psi_l^{(0)}(\nu,r)=\frac{I_{l+1}(\kappa r)}{r} \kma
\ee
and we have
\be                                                  \label{h}
\psi_l(\nu,r)=\left[1+h_l(\nu,r)\right]\frac{I_{l+1}(\kappa r)}{r}\pkt
\ee
Then by the theorem (\ref{th}), the ratio of determinants
(\ref{Dl}) can be expressed as
\be
J_l(\nu ) = 1+h_l(\nu,\infty) \pkt
\ee
In terms of the $h$ function the first equation (\ref{eth})
reads
\be                                                 \label{eh}
\left\{\frac{d^2}{dr^2}+\left[2\kappa\frac{I_{l+1}'(\kappa r)}{I_{l+1}(\kappa r)}
+\frac{1}{ r} \right]\frac{d}{dr}\right\}h_l(\nu,r)
=V(r)\left[1+h_l(\nu,r)\right]\kma
\ee
where $I_{l+1}'(\kappa r)\equiv dI_{l+1}(\kappa r)/d(\kappa r)$.

In what follows it would be convenient to consider the perturbation expansion
\be                                                  \label{hk}
h_l(\nu,r)=\sum_{k=1}^\infty h^{(k)}_l(\nu,r)
\ee
in powers of the potential $V(r)$. This assumes an analogous
expansion for the ratios $J_l(\nu )$ in the sense that
$J_l^{(k)}(\nu)=h_l^{(k)}(\nu,\infty)$.
The $k$-order contribution $h_l^{(k)}$ obeys an equation
\be                                                 \label{ehk}
\left\{\frac{d^2}{ dr^2}+\left[2\kappa\frac{I_{l+1}'(\kappa r)}
{I_{l+1}(\kappa r)}+\frac{1}{r} \right]
\frac{d}{ dr} \right\} h_l^{(k)}(\nu,r)=V(r)h_l^{(k-1)}(\nu,r)\kma
\ee
where we defined $h_l^{(0)}\equiv 1$.
Since Eq.~(\ref{ehk}) is linear differential equation it holds also
for linear combinations of $h_l^{(k)}$. It is convenient to introduce
notation $h_l^{\overline{(k)}}=\sum_{q=k}^\infty h_l^{(q)}$.
In this notation $h_l=h_l^{\overline{(1)}}$.
A Green function that gives the solution to
equation (\ref{ehk}) in the form
\be                                               \label{defG}
h_l^{\overline{(k)}}(r)
=-\int_0^\infty d\tir \tir G_l(r,\tir)V(\tir)h_l^{\overline{(k-1)}}(\tir)
\ee
with the right boundary condition at $r=0$ reads
\be                                                   \label{G}
G_l(r,\tir)=  \frac{I_{l+1}(\kappa \tir)}{I_{l+1}(\kappa r)}
\left[ I_{l+1}(\kappa r_<)K_{l+1}(\kappa r_>)
-I_{l+1}(\kappa r)K_{l+1}(\kappa \tir) \right] \kma
\ee
where $r_<=\min \{r,\tir\}$, $r_>=\max \{r,\tir\}$.

The first term on the right-hand side of Eq.(\ref{G})
does not contribute to $h_l^{(k)}(\infty)$. The Green function (\ref{G})
gives rise to connected graphs as well as disconnected ones.
The latter are cancelled in
$\ln (1+h_l(\infty))$ whose expansion in $k$-order connected graphs
$J^{(k)}_{l~con}(\nu)$ reads
\be                                              \label{Jconn}
\ln J_l(\nu ) = \ln (1+h_l(\nu, \infty))
=\sum_{k=1}^\infty \frac{(-1)^{k+1}}{k}J^{(k)}_{l~con}(\nu)\pkt
\ee

This formula is analogous to the expansion of the full functional
determinant in terms of Feynman diagrams
\be                                              \label{lnD}
\ln {\cal D}
=\sum_{k=1}^\infty \frac{(-1)^{k+1}}{k} A^{(k)} \kma
\ee
where $A^{(k)}$ is the one-loop  Feynman graph of order $k$
in the external potential $V(r)$.

Indeed, it is obvious from Eq.(\ref{defG}) that $h_l^{(k)}$ and,
therefore, $J^{(k)}_{l~con}$
are of the order $V^k$. Since the expansion
of $\ln {\cal D}$ in powers of $V$ is unique, we conclude that
\be                                              \label{AandJ}
A^{(k)}= \sum_{l=0}^\infty (l+1)^2 J^{(k)}_{l~con} \pkt
\ee

\subsection{Calculation of ${\cal D}^{\overline{(3)}}$}          \label{s2}
\par

Making use of a uniform asymptotic expansion of the modified Bessel
functions in (\ref{G}) one can check that
that $J^{(k)}_{l~con}\sim 1/l^{2k-1}$ as $l\rightarrow \infty $.
That results in the expected quadratic and logarithmic ultraviolet
divergences in $\ln {\cal D}$ due to the contribution of
$J^{(1)}_{l~con}$ and $J^{(2)}_{l~con}$.
Our strategy is to compute {\it analytically} the first two terms
in the sum Eq.~(\ref{lnD}) and to add {\it numerically}
computed $\ln {\cal D}^{\overline{(3)}}$, which is the sum
without first and second order diagrams $A^{(1)}$ and $A^{(2)}$.
It reads explicitly
\be                                                \label{sum}
\ln{\cal D}^{\overline{(3)}} = \sum_{l=0}^\infty  (l+1)^2
\left( \ln J_l(\nu ) \right) ^{\overline{(3)}} \kma
\ee
where
\be                                               \label{J3b}
\left( \ln J_l(\nu ) \right) ^{\overline{(3)}} =
\ln\left(1+h_l(\infty)\right)- h_l^{(1)}(\infty)-\left[h_l^{(2)}(\infty)
- \frac{1}{2} \left(h_l^{(1)}(\infty)\right)^2 \right] \pkt
\ee
The terms in square brackets here correspond to the fish diagram
$J^{(2)}_{l~con}$. Since all contributions to
$\ln {\cal D}^{\overline{(3)}}$ are
ultraviolet finite, we need no regularization in computing them.
The divergent contributions of the first and second order in $V$
will be considered in Sec. \ref{ren}.

In order to avoid numerical subtraction that might be delicate
we re-write the term (\ref{J3b}) to
be summed up on the right-hand side (\ref{sum}) in the form
\begin{eqnarray}                                            \label{hren3}
\displaystyle
\left( \ln J_l(\nu ) \right) ^{\overline{(3)}}
&=&\left[\ln(1+h_l(\infty))
-h_l(\infty) + \frac{1}{2} h_l(\infty)^2 \right]  \nonumber   \\
\displaystyle
&&+~ h_l^{\overline{(3)}}(\infty)
-\frac{1}{2} h_l^{\overline{(2)}}(\infty)
\left(h_l(\infty) + h_l^{(1)}(\infty) \right) \; .
\end{eqnarray}
Each of the three terms on the r.h.s. is now manifestly of order $V^3$.
The subtraction done in the square bracket is exact enough when
the logarithm is calculated with double precision.
We have determined $h_l(r)$ as solutions of Eq.(\ref{eh})
and $h_l^{(1)}(r)$, $h_l^{\overline{(2)}}(r)$ and $h_l^{\overline{(3)}}(r)$
as those of Eq.(\ref{ehk})
using Runge-Kutta-Nystr\"om integration method \cite{zurmuel}.
Of course we cannot integrate the differential equations until
$r=\infty$. In fact we have integrated it up to the maximal
value for which we know the profile $\phi(r)$, and therefore $V(r)$.
This value is such, that the classical field has well reached its vacuum 
expectation value, and therefore $V(r)$ has become zero. This is 
the condition under which we can impose the asymptotic boundary condition 
for the classical profile.
For such values the functions $h^{(k)}_l(r)$ have already become constant;
indeed for $V(r)=0$ they  have the exact form
$a+b K_{l+1}(\kappa r)/I_{l+1}(\kappa r)$ and the second part decreases 
exponentially for $r >> 1/\kappa$. In praxi we used values of $R$
up to  $R_{\max}=m r_{\max}\simeq 20-30$.

We have neglected till now the existence of the negative mode
$\omega_{0}^2<0$ for $l=0$ and four zero modes $\omega_{1}^2=0$
with $l=1$.
The former results in negative value of $J_0(\nu )=1+h_0(\nu,\infty )$
at $\nu=0$. According to Eq.(\ref{rate}) one has to replace
$\omega_{0}^2$ by $|\omega_{0}^2|$. That implies
taking the absolute value of $J_0(0)$ in Eq.(\ref{sum});
indeed $J_0(0)$ is found to be negative.

The translational zero modes manifest themselves by the 
vanishing of $\omega^2_{10}=0$, the lowest radial excitation
in the $l=1$ channel with degeneracy $(l+1)^2=4$, and thereby by
the vanishing of $J_1(\nu )$ at $\nu =0$, see Eq.\eqn{Dl}.
This represents a good check for both the classical solution
and for the integration of the partial waves.
The factor $\nu^2$ has to be removed according to the definition
of $\det'$. So in the $l=1$ contribution we have to replace $J_1(0)$ by
\be                                                 \label{h'}
\lim_{\nu\to  0}\frac{J_1(\nu)}{\nu^2}=
\frac{d J_1(\nu)}{d\nu^2}=\frac{d}{d(\nu^2)}
h_1(\nu,\infty) { \mid}_{\nu=0} \pkt
\ee
Notice that replacement Eq.~(\ref{h'}) introduces a dimension into the
functional determinant. Thereby the units used for $\nu$ become
the units of the transition rate. Here we have used the
scale $m$ throughout, see  Eqs. (\ref{rate1}) and (\ref{seff1}).

Our next step is performing summation over $l$ in Eq.(\ref{sum}).
For small bounces ($\alpha \lesssim 0.8$) we have found good agreement with
the expected behavior, namely
\be                                                \label{asl5}
\left( \ln J_l(\nu ) \right)^{\overline{(3)}}\propto \frac{1}{(l+1)^5} \pkt
\ee
So, the summation has been done by cutting
the sum at some value $l_{max}$
and adding the rest sum from $l_{max}+1$ to $\infty$ of terms fitted with
\be                                               \label{asy}
\ln J_l^{\overline{(3)}} \approx
\frac{a}{(l+1)^5}+\frac{b}{(l+1)^6} +\frac{c}{(l+1)^7}\pkt
\ee
The summation was stopped when increasing of
$l_{max}$ by unity did not change the result within some
given accuracy $\delta$. The required accuracy was decreased for higher
$\alpha$. The problem is that the convergence becomes worse as we get
closer to $\alpha=1 $. This is related to the fact
that the asymptotic behavior (\ref{asl5})
sets in only when $l\gg m r_{\rm eff}$,
where $r_{\rm eff}$ is the characteristic size of the bounce.
It is of order $1/m$ at small values of $\alpha$ and can be
estimated as $1/(1-\alpha)m$ near the thin-wall
limit, $\alpha\rightarrow 1$. As the maximal value of the angular
momentum that we have used is $l=25$, our computations cease to be
reliable beyond $\alpha \simeq 0.95$. The value of $\delta$ was
about $10^{-5}$ for small
bounces, and of order of $10^{-3}$ for $\alpha > 0.85$.
As we will see below, for larger values of $\alpha$ the effective
action is well approximated by the leading terms of a gradient
expansion.


\subsection{Perturbative contribution and renormalization}       \label{ren}
\par

We have described in the previous subsection  the computation of the
finite part $\ln {\cal D}^{\overline{(3)}}$ which is the sum
of all one-loop diagrams of the third order and higher,
\be                                              \label{lnD3b}
\ln {\cal D}^{\overline{(3)}}
=\sum_{k=3}^\infty \frac{(-1)^{k+1}}{k} A^{(k)}~~~.
\ee
We now have to discuss the leading divergent contributions
$  A^{(1)}$ and $A^{(2)}$. These are computed as ordinary Feynman graphs.
Using dimensional regularization we have
\be
 A^{(1)}=\int\frac{d^{4-\epsilon}k}{(2\pi)^{4-\epsilon}}
\frac{\tilde V(0)}{k^2+m^2}
\ee
where we have introduced the Fourier transform of the potential
\be
\tilde V(k)=\int d^4 x V(x)e^{-ikx} \pkt
\ee
We obtain
\be
A^{(1)}=-\frac{m^2}{16 \pi^2}\left[\frac{2}{\epsilon}-\gamma_E
+\ln 4\pi+\ln \frac{\mu^2}{m^2}+1\right]\int d^4 x V(x) \kma
\ee
where $\mu$ is the usual dimensional regularization parameter.
We choose it to be equal to $m$.
Then using the $\overline{MS}$ scheme we just retain the last
contribution in the bracket (see e.g. \cite{psbook}, p. 377).
Thus, the finite part of $A^{(1)}$ is
\be
A^{(1)}_{fin}=-\frac{1}{8} \int_{0}^{\infty} R^3 dR~V(R) \pkt
\ee

The second order terms takes the form
\be
 A^{(2)}=
\int\frac{d^{4-\epsilon}q}{(2\pi)^{4-\epsilon}} \biggl|\tilde V(q)\biggr|^2
\int\frac{d^{4-\epsilon}k}{(2\pi)^{4-\epsilon}}
\frac{1}{(k^2+m^2)[(k+q)^2+m^2]} \pkt
\ee
We obtain
\bea
 A^{(2)}&=& \frac{1}{16 \pi^2}\left[\frac{2}{\epsilon}-\gamma_E
+\ln 4\pi+\ln \frac{\mu^2}{m^2}\right]\int d^4 x (V(x))^2
\\ \nonumber
&& + \frac{1}{128 \pi^4}\int q^3 dq \biggl|\tilde V(q)\biggr|^2 \left[2-
\frac{\sqrt{q^2+4m^2}}{q}\ln\frac{\sqrt{q^2+4m^2}+q}{\sqrt{q^2+4m^2}-q}
\right] \pkt
\eea
Again the  $\overline{MS}$ scheme corresponds to omitting the
first term on the right hand side and for the finite part of $A^{(2)}$
we find
\be
A^{(2)}_{fin}=\frac{1}{128 \pi^4}\int_{0}^{\infty}
Q^3 dQ \biggl|\tilde V(Q)\biggr|^2
\left[ 2- \sqrt{Q^2+4}~\ln\frac{\sqrt{Q^2+4}+Q}{\sqrt{Q^2+4}-Q}\right] \kma
\ee
with $Q=q/m$ being the dimensionless momenta.
For the numerical evaluation of $A^{(2)}$ we have to compute the Fourier transform
of the external potential which is known numerically, the remaining  computation is straightforward.

\setcounter{equation}{0}
\section{Numerical results}                                \label{num}
\par

To summarize we represented the false vacuum decay rate per unit  time
per unit volume as
\be                                               \label{rate1}
\gamma = m^4 \left(\frac{S_{cl}[\varphi]}{2\pi}\right)^2
\e^{-S_{cl}[\varphi] -S^{eff}_{1-loop}[\varphi]} \kma
\ee
where
\be \label{seff1}
S^{eff}_{1-loop}[\varphi]= \frac{1}{2}\ln |m^8 \cald[\varphi]|
=S^{eff}_{\rm 1-loop, p}+S^{eff}_{\rm 1-loop, n.p.} \kma
\ee
with perturbative
\be
S^{eff}_{\rm 1-loop, p} =\frac{1}{2}(A^{(1)}_{fin}- \frac{1}{2} A^{(2)}_{fin})
\ee
and non perturbative
\be
S^{eff}_{\rm 1-loop, n.p.}=
\frac{1}{2} \sum_{k=3}^\infty \frac{(-1)^{k+1}}{k} A^{(k)}
=\frac{1}{2} \ln |{\cal D}^{\overline{(3)}}|
\ee
contributions.

It is useful to introduce the quantity $G$,
\be
G(\alpha,\beta)=S^{eff}_{\rm 1-loop}[\varphi] / S_{cl}[\varphi_b] \kma
\ee
which indicates how big quantum corrections are. Since classical action,
Eq.~(\ref{scl}), depends linearly on parameter $\beta$ we have
$G(\alpha,\beta)=G(\alpha,1)/\beta$.
Numerical calculation shows that that $G(\alpha, 1)$ varies from 0.0367 to
0.0448 as we vary $\alpha $ from 0 to 0.95, with shallow minimum
$G_{min}\approx 0.033$ at $\alpha$ about 0.6 (see Fig.~\ref{Gplot}).
Fig.~\ref{Gplot} suggests that $G(1,1)\approx 0.05$,
which means that for sufficiently big values of $\beta$, namely $\beta > 0.1$,
the quantum corrections to the classical action are small (less then $50\%$)
for all values of $\alpha$.

The corrections to the {\em transition rate} are given directly by
a factor $\exp(-S^{eff}_{\rm 1-loop})$, so even if the classical transition 
rate
is sizeable, as it happens for small $\beta$, the quantum corrections suppress
the decay of the false vacuum by factors $\exp (-3.3)$ at $\alpha=0$ 
and $\exp(-291)$ at $\alpha=.9$.

Note that the main contribution to the effective action for {\it all}
$\alpha$ is coming from the $A^{(1)}_{fin}$, (comp. Tab.~1, Tab.~2).
For small $\alpha$ perturbative contribution is almost $100\%$ of total
one loop effective action, (comp. Fig.~\ref{srg_sef}).

In the limit $\alpha \to 1$ the leading terms of the gradient expansion
(Appendix B) gives dominant contribution to the one loop effective action.
Already for $\alpha = 0.8$ the sum of leading gradient terms
\be
S^{eff}_{\rm grad, 0+2}=
S^{eff}_{{\rm grad}, 0}+S^{eff}_{{\rm grad}, 2}
\ee
approximates the $1-$loop effective action $S^{eff}_{\rm 1-loop}$
within 20$\%$.
So the gradient expansion reproduces well the behavior of the
one-loop effective action when $\alpha \to 1$, see Fig. \ref{srg_sef}. 
As the numerical
procedure described in the main part of this paper becomes 
precarious for $\alpha \gtrsim  0.9$ this expansion complements the computation
of the transition rate in this region.

As it is well known there is exactly one negative mode in the spectrum
of fluctuations about the bounce.
Its energy is plotted vs $\alpha$ in Fig. \ref{Omega}.

In the present paper we used dimensional regularization and
have chosen the parameter $\mu^2$, which can be understood as parameterizing
a sequence of possible renormalization conditions, to be equal $m^2$.
Choosing $\mu^2$ differently would result in the following corrections to
$A^{(1)}_{fin}$ and $A^{(2)}_{fin}$
\bea
A^{(1)}_{fin}\to (1+ \ln \frac{\mu^2}{m^2}) A^{(1)}_{fin} \nn \\
A^{(2)}_{fin} \to A^{(2)}_{fin}+ a^{(2)} \ln \frac{\mu^2}{m^2}
\eea
where $a^{(2)}$ is the  following integral
\be
a^{(2)}=  \frac{1}{8} \int_{0}^{\infty} R^3 dR (V(R))^2
\ee
evaluated at the bounce solution. Numerical values for $A^{(1)}_{fin},
A^{(2)}_{fin}$ and $a^{(2)}$ for different
values of $\alpha$ are collected in Tab.~2.
With the present choice of $\mu^2$ the perturbative terms represent the most
important contributions to the effective action (see above),
this means at the same time that a modification of the regularization
and renormalization procedures can result in
large changes in the one-loop effective action.

\setcounter{equation}{0}
\section{Discussion and Conclusion}   \label{concl}
\par

In the present paper we applied previously developed technique for
evaluations of functional determinants and calculated quantum
corrections to the tunnelling transitions in 3+1 dimensional model
of one self-interacting scalar field.

In the present toy model decay rate is vanishingly small.
The sign of quantum corrections is such that it decreases false vacuum
decay rate.
The corrections can be thought as originating from the particles creation
during the phase transitions.
The created particles take energy from the tunnelling field and therefore
decrease tunnelling probability.
Analytical estimations show that particle creation is typically weak
in the thin-wall approximation \cite{rubakov84}.
In the present paper it was found that the quantum corrections are even
smaller away from the thin-wall case (compare Fig.~\ref{Gplot}),
which assumes that particle creation for $\beta > 0.1$
is weak for all values of the coupling constant $\alpha$.
On the other hand for $\beta < 0.1$ the quantum corrections dominate,
which means that in this regime
one should look for a bounce solution taking into account
the full effective action in the one-loop approximation
\cite{surig97,levkov02}.

Corrections to the false vacuum decay in a similar model in (3+1)
dimensional theory in {\it the thin wall approximation} with the
heat kernel expansion technique were calculated in \cite{kr85},
but it is not straightforward to compare our results since
we use a different renormalization scheme and and a different
parametrization of the potential.
Powerful techniques for analytic calculations of the pre-factor
using different approximations were developed in \cite{mr99,stw99,mst00},
but we cannot compare our results directly, since these calculations are
within 3d theory.

The technique described here can be applied to tunnelling transitions
in more realistic theories in 4 dimensions.

\section*{Acknowledgments}

G.L. if thankful to Theory Group of the University of Dortmund for kind hospitality
during his visit to Dortmund, where this work started
and to the theory groups of Max-Planck-Institute for Physics
and Max-Planck-Institute for Gravitational Physics
for stimulating and fruitful atmosphere
during his visits to Munich and Golm, where part of the work was done.
Work of G.L. was partly supported by Grant of the Georgian Academy of Sciences.

\appendix
\setcounter{section}{1}
\section*{Appendix A. The thin-wall approximation}

In the limit $\alpha \to 1$ so called thin-wall case is realized.
This is when energy density difference $\epsilon$ between two vacuums
\be
\epsilon = U(\Phi_{-})- U(\Phi_{+})\kma
\ee
is small (compare to the hight of the barrier).
In this case potential
Eq. (\ref{pot}) can be represented as
\be                                             \label{pottw}
U(\Phi)= U_{0}(\Phi)+O(\epsilon)\kma
\ee
where symmetric part, $U_{0}$, in our case is
\be                                             \label{potkink}
U_{0}(\Phi)=\frac{1}{8}\Phi^2 (2-{\Phi})^2 \kma
\ee
and
\be
\epsilon= 2(1-\alpha) \kma
\ee

In the thin-wall approximation the  radius $\bar{R}$ of the bounce and
the Euclidean action $S_{cl}$ are given analytically \cite{Co77,Co79} as
\be
\bar{R} =\frac{3 S_{1}}{\epsilon}, \qquad
\tilde{S}^{tw}_{cl}=\frac{27 \pi^2 S_{1}^4}{2 \epsilon^3} \kma
\ee
where
\be
S_{1}= 2 \int_{-\infty}^{\infty} dR~~U_{0}(\Phi_{k}) \kma
\ee
is the action of the one-dimensional kink solution corresponding to
degenerate potential $U_{0}$ with the equal minima.
For our choice of the potential, Eq.~(\ref{potkink}), the kink solutions is
\be
\Phi_{k} = \frac{2}{1+\e^{(R-\bar{R})}}\pkt
\ee
One finds that $S_1 = 2/3$ and correspondingly
\be
\bar{R} = \frac{1}{1-\alpha},
\qquad \tilde{S}^{tw}_{cl} = \frac{\pi^2}{3 (1-\alpha)^3}\pkt
\ee


\setcounter{section}{2}
\setcounter{equation}{0}
\section*{Appendix B. The leading terms of the gradient expansion}

We want to derive an approximation to the effective action of a scalar field on the background
of a bounce solution. The strategy is to expand first the effective action
with respect to external vertices, and to expand in a second step the resulting Feynman amplitudes
with respect to the external momenta. This approach is fairly standard, and has been
used, e.g., in Ref. \cite{Caro:1993fs}. We note that we will retain
all powers in the external vertices; such a summation was found to
yield a very good approximation for the sphaleron determinant
\cite{Carson:1990jm,baju93}, see Fig. 1 in the second 
entry of Ref. \cite{baju93}.
We have to compute the trace log or log det of
a generalized Euclidean Klein-Gordon operator $\Delta_4 + U''(\phi)$
where $\Delta_4$ is the four-dimensional Laplace operator. Formally
\be
\left[\ln \cald\right]=\ln \left[\frac{-\Delta_4+U''(\phi)}{-\Delta_4+U''(0)}\right]
\pkt
\ee
We introduce a potential $V(x)$ via
\be
U''(\phi(x))=m^2+V(x) \; ; \;\;\; U''(0)=m^2 \pkt
\ee
For the bounce the potential depends only on $r=|x|$ but we will not use this
now. The logarithm can be expanded with respect to the potential $V(x)$.
We write
\bea \nonumber
\left[\ln \cald\right]&=&\ln \left[\frac{-\Delta_4+m^2+V(x)}{-\Delta_4+m^2}\right]
\\
&=&\ln \left[(-\Delta_4+m^2)^{-1}(\Delta_4+m^2+V(x))\right]
\\ \nonumber
&=&\ln \left[1 + (-\Delta_4+m^2)^{-1}V(x)\right]
\\ \nonumber
&=&\sum_{N=1}^\infty\frac{(-1)^{N+1}}{N}\left[(-\Delta_4+m^2)^{-1}V(x)\right]^N
\kma
\eea
and the effective action is given by
\be
S^{\rm eff}=\sum_{N=1}^\infty\frac{(-1)^{N+1}}{2N}
\tr\left[(-\Delta_4+m^2)^{-1}V(x)\right]^N \pkt
\ee
We introduce the Fourier transform
\be
\tilde V(q)=\int e^{-iq\cdot x}V(x)d^4x
\pkt\ee
The individual terms in the expansion of the effective action have the
form of Feynman
diagrams with external sources $V(q_j)$ with $j=1\dots k$. The momentum that has
flown into the line $l$ is
\be
Q_l=\sum_{j=1}^l q_j
\kma
\ee
of course the total momentum must be zero,i.e., $Q_N=0$.
With these notations we can write the $N$th term in the effective action,
omitting the factor $(-1)^{N+1}/2N$ as
\be
A_N = \int \frac{d^4p}{(2\pi)^4}\prod_{j=1}^N\left[\int \frac{d^4q_j}{(2\pi)^4}
\tilde V(q_j)\right]\prod_{l=1}^N\left[\frac{1}{(p+Q_l)^2+m^2}\right](2\pi)^4
\delta(Q_N)
\pkt
\ee
The four-momentum delta function arises from taking the trace.
We obtain a gradient expansion by expanding the
denominators $(p+Q_l)^2+m^2$ with respect to the momenta $Q_l$.
The leading term is of course
\bea \nonumber
A_{N,0}&=& \int \frac{d^4p}{(2\pi)^4}\left[\frac{1}{p^2+m^2}\right]^N
\prod_{j=1}^N\left[\int \frac{d^4q_j}{(2\pi)^4}
\tilde V(q_j)\right](2\pi)^4\delta(Q_N)
\\
&=&\int \frac{d^4p}{(2\pi)^4}\left[\frac{1}{p^2+m^2}\right]^N
\int d^4x \left[V(x)\right]^N \pkt
\eea
The zero-gradient contribution to the effective action is obtained by
resuming this series; one finds
\be
S^{\rm eff}_{{\rm grad}, 0}=\frac{1}{2}\int d^4x \int\frac{d^4p}{(2\pi)^4}
\ln \left\{\frac{p^2+U''(\phi)}{p^2+U''(0)}\right\} \equiv
\frac{1}{2}\int d^4x K^{(4)} \pkt
\ee
Of course this integral has to be regularized, e.g., via dimensional
regularization. The divergences come form the terms with $N=1$ and $N=2$,
which are standard divergent one loop integrals.

We find
\beast
 K^{(D)}&=& \frac{2 \pi^{D/2}}{\Gamma\left(\frac{D}{2}\right)}
\int \frac{dp\,p^{D-1}}{(2\pi)^D}
\ln \left[\frac{p^2+U''(\phi)}{p^2+U''(0)}\right]\\
&=&\frac{2}{\Gamma\left(\frac{D}{2}\right)(4\pi)^{D/2}}
\left\{\frac{1}{D}p^D\ln \left[\frac{p^2+U''(\phi)}{p^2+U''(0)}\right]
\biggl|_{p=0}^\infty\right.
\\ && \left.\hspace{10mm}-\frac{2}{D}\int dp\,p^{D+1}\left[
\frac{1}{p^2+U''(\phi)}-\frac{1}{p^2+U''(0)}\right]\right\}
\pkt
\eeast
The first term in the parenthesis vanishes for $0<D<2$ and
is defined to vanish in general by analytic continuation.
The second term can be rewritten as
\beast
&&\frac{-2}{D\Gamma\left(\frac{D}{2}\right)(4\pi)^{D/2}}
\left\{[U''(\phi)]^{D/2}-[U''(0)]^{D/2}\right\}\int_0^\infty
dx \frac{x^{D+1}}{x^2+1}\\
&&= \frac{-2}{D\Gamma\left(\frac{D}{2}\right)(4\pi)^{D/2}}
\frac{\Gamma(D/2+1)\Gamma(-D/2)}{2\Gamma(1)}
\left\{[U''(\phi)]^{D/2}-[U''(0)]^{D/2}\right\}
\\
&&=-\frac{\Gamma(-D/2)}{(4\pi)^{D/2}}
\left\{[U''(\phi)]^{D/2}-[U''(0)]^{D/2}\right\}
 \pkt
 \eeast
Now set $D=4-\epsilon$ and use
\beast
\Gamma\left(-\frac{D}{2}\right)&=&
\frac{1}{(-2+\epsilon/2)(-1+\epsilon/2)}
\Gamma\left(\frac{\epsilon}{2}\right) \\
&=& \frac{1}{2}\left\{\frac{2}{\epsilon}-\gamma_E +\frac{3}{2}\right\}
\eeast
to obtain
\beast
K^{(4-\epsilon)} &=&\frac{-1}{32\pi^2}
\left[\frac{2}{\epsilon}-\gamma_E +\ln 4\pi+\frac{3}{2}\right]
\\&&\times\left\{(m^2+V(r))^2\left[1-\frac{\epsilon}{2}
\ln \frac{m^2+V(r)}{\mu^2})\right]
-m^4\left[1-\frac{\epsilon}{2}\ln \frac{m^2}{\mu^2}\right]\right\} \pkt
\eeast
Using $\overline{MS}$ subtraction
we get
\bea
K^{(4)}&=&
\frac{-1}{32\pi^2}\left\{\frac{3}{2}\left[2m^2 V(r)+V^2(r)\right]
\right.\\\nonumber
&&\left. -\left(m^2+V(r)\right)^2\ln\frac{m^2+V(r)}{\mu^2}
+m^4\ln \frac{m^2}{\mu^2} \right\} \pkt
\eea
Integrating over 4d Euclidean space we finally obtain
\bea
S^{eff}_{{\rm grad}, 0}&=&
\frac{1}{32} \int^{\infty}_{0} R^3 dR \left[
\left(1+V(R)\right)^2\ln \frac{1+V(R)}{\tilde{\mu}^2} \right.  \nonumber \\
&-& \left. \frac{3}{2}\left(2V(R)+V^2(R)\right)+ \ln \tilde{\mu}^2
\right] \kma
\eea
with $\tilde{\mu}=\mu/m$.

Let us now consider the one- and two-gradient contributions.
We expand the denominators up to second order in the gradients,
i.e., in the momenta $Q_j$. We obtain
\bea
\nonumber
\Pi_N &\equiv& \prod_{l=1}^N\left[\frac{1}{(p+Q_l)^2+m^2}\right]
=\frac{1}{(p^2+m^2)^N}-\frac{1}{(p^2+m^2)^{N+1}}\sum_{j=1}^N 2p\cdot Q_j
\\ \nonumber
&&-\frac{1}{(p^2+m^2)^{N+1}}\sum_{j=1}^N Q_j^2+
\frac{1}{(p^2+m^2)^{N+2}}\sum_{j=1}^{N-1}\sum_{k=j+1}^N
4 (p\cdot Q_j)(p\cdot Q_k) \\
&&+\frac{1}{(p^2+m^2)^{N+2}}\sum_{j=1}^N 4(p\cdot Q_j)^2 + O(Q^3) pkt
\eea
Under $O(4)-$symmetric integration $4 p_\mu p_\nu \simeq p^2 \delta_{\mu\nu}$,
and $p_\mu \simeq 0$. So the one-gradient term vanishes and the complete
two-gradient contribution becomes
\bea \nonumber
\Pi_{N,2}&=&\frac{1}{(p^2+m^2)^{N+2}}
\left[-(p^2+m^2)\sum_j Q_j^2+4p_\mu p_\nu \sum_{k>j}Q_{j\mu}Q_{k\nu}
+4 p_\mu p_\nu \sum_j Q_{j\mu}Q_{j\nu}\right]
\\ &\simeq&\frac{1}{(p^2+m^2)^{N+2}}
 \left[p^2 \sum_{k > j}Q_j\cdot Q_k
-m^2 \sum_j Q_j^2\right] \pkt
\eea
We now have to rewrite this in terms of the momenta $q_j$ that represent
the gradients on the functions $V(q_j)$. After having used the fact that
$\Pi_2$ appears under the integral over $d^4p$ we will now use the fact
that it appears under the product of integrals  $\int d^4q_j V(q_j)$ which
implies permutation symmetry in the indices $j$.
So if we expand the products $Q_j\cdot Q_k$ and $Q_j^2$ we will
encounter just two kinds of terms: products $q_l\cdot q_m$
with $l\neq m$ and squares $q_l^2$, which may be replaced by
$q_1\cdot q_2$ and by $q_1^2$, respectively. We have to do some
combinatorics in order to find
\bea
\sum_j Q_j^2 &\simeq& \frac{(N-1)N(N+1)}{3}q_1\cdot q_2 +
\frac{N(N+1)}{2} q_1^2
\\ \nonumber \sum_{k > j}Q_j\cdot Q_k
&\simeq& \frac{(N-1)N(N+1)(3N-2)}{24}q_1\cdot q_2
\\
&&+ \frac{(N-1)N(N+1)}{6}q_1^2
\pkt
\eea
Now we may use momentum conservation to rewrite
\be
q_1^2 =-q_1\cdot (q_2 + \dots + q_N)\simeq -(N-1)q_1\cdot q_2
\ee
 so that
\bea
\sum_j Q_j^2 &\simeq& -\frac{(N-1)N(N+1)}{6}q_1\cdot q_2
\\
\sum_{k>j} Q_j\cdot Q_k &\simeq&
-\frac{(N-2)(N-1)N(N+1)}{24}q_1\cdot q_2
\eea
and
\be
\Pi_{N,2} \simeq \frac{1}{(p^2+m^2)^{N+2}}\frac{(N-1)N(N+1)}{24}q_1\cdot q_2
\left[-(N-2) p^2 +4 m^2\right] \pkt
\ee
The momentum integrals are
\bea
\int \frac{d^4p}{(2\pi)^4}\frac{p^2}{(p^2+m^2)^{N+2}}
&=&\frac{1}{16 \pi^2} m^{2-2N}\frac{2}{(N-1)N(N+1)}
\\
\int \frac{d^4p}{(2\pi)^4}\frac{m^2}{(p^2+m^2)^{N+2}}
&=&\frac{1}{16 \pi^2} m^{2-2N}\frac{1}{N(N+1)}
\eea
and, therefore,
\be
\int \frac{d^4p}{(2\pi)^4}\Pi_{N,2}=
q_1\cdot q_2 \frac{1}{16 \pi^2}m^{2-2N}\frac{N}{12}
\pkt
\ee
The momenta are converted into gradients; so we finally obtain
as the expansion terms of the two-gradient part of the effective action
\be
A_{N,2}= - \frac{1}{16 \pi^2}\int d^4 x
\left[\frac{V(x)}{m^2}\right]^{N-2}\frac{N}{12m^2}
\left(\nabla V(x)\right)^2 \pkt
\ee
The term $A_{1,2}$ is zero. The sum over all terms yields
\be \label{grad2}
S^{\rm eff}_{grad, 2}=
\frac{1}{32 \pi^2}\int d^4x \frac{1}{m^2+V(x)}
\frac{1}{12}\left(\nabla V(x)\right)^2 \kma
\ee
or finally in dimensionless variables
\be
S^{\rm eff}_{grad, 2}=
\frac{1}{192}\int_0^\infty R^3 dR \frac{1}{1+V(R)}\left(V'(R)\right)^2 \pkt
\ee

An alternative derivation starts with a technical step that frees us from the
denominator $1/N$. We take the derivative of the effective action with respect to
$m^2$, a step that we can revert later on.
We then obtain, using the cyclic property of the trace,
\bea \nonumber
\calg&\equiv& \frac{dS^{\rm eff}}{dm^2}
\\
&=&\sum_{N=0}^\infty \frac{(-1)^N}{2}
\tr\left\{\left[(-\Delta_4+m^2)^{-1}V(x)\right]^N(-\Delta_4+m^2)^{-1}\right\}
\\ \nonumber
&=&\frac{1}{2}\sum_{N=0}^\infty B_N
\pkt\eea
We note that we have included the $N=0$ term,
which can be removed later on if necessary. So we have
arrived at the trace of the exact Green function in the
external field. The terms $B_N$ have the form
\bea \nonumber
B_N &=&(-1)^N \int \frac{d^4p}{(2\pi)^4}
\prod_{j=1}^N\left[\int \frac{d^4q_j}{(2\pi)^4}\right]
\frac{1}{p^2+m^2}\tilde V(q_1)\frac{1}{(p+Q_1)^2+m^2}
\\ && \times \tilde V(q_2)
\frac{1}{(p+Q_2)^2+m^2}\tilde V(q_3)
\dots
\\ \nonumber && \times
\tilde V(q_N)\frac{1}{(p+Q_N)^2+m^2} (2\pi)^4
\delta(Q_N)
\pkt
\eea
Assume we have expanded the fraction $1/[(p+Q_k)^2+m^2]$
to first order in $2p\cdot Q_k+Q_k^2$, yielding a factor
\be
\frac{1}{p^2+m^2}\left[-2p\cdot Q_k-Q_k^2\right]\frac{1}{p^2+m^2}
\ee
at the $k^{\rm th}$ place in the product of propagators and vertices,
in other words we have obtained an insertion
of $-2p\cdot Q_k-Q_k^2$. Consider the part of the product
to the right of this insertion.
We rewrite it as
\bea\nonumber
&\prod_{j=k+1}^N&\left[\int \frac{d^4q_j}{(2\pi)^4}\right]
\left[-2p\cdot Q_k-Q_k^2\right]\frac{1}{p^2+m^2}
\\ \label{reststring}
\times &\prod_{j=k+1}^N& \left[\int d^4x_jV(x_j)\frac{e^{-iq_j\cdot x_j}}
{p^2+m^2}\right] (2\pi)^4\delta(Q_k+q_{k+1}+\dots+q_N) \pkt
\eea
We furthermore rewrite the delta function as
\be
(2\pi)^4\delta(Q_k+q_{k+1}+\dots+q_N)
=\int d^4x e^{i(Q_k+q_{k+1}+\dots+q_N)\cdot x} \pkt
\ee
Inserting this in \eqn{reststring} we can carry out the integrations
over  the $q_j$ and the $x_j$ to obtain
\be
\int d^4x e^{iQ_k\cdot x}\left[-2p\cdot Q_k-Q_k^2\right]
\frac{1}{p^2+m^2}
\prod_{j=k+1}^N\left[V(x)\frac{1}{p^2+m^2}\right]
\pkt\ee
Now the $Q_{k,\mu}$ in $2p\cdot Q_k+Q_k^2$ can be written as
$-i\partial/\partial x_\mu =-i \partial_\mu$ on the exponential.
Integrating by parts they can be written as $i\partial_\mu$
acting on the product to their right. So the whole string
to the right of the insertion can be written as
\be
\int d^4x e^{iQ_k\cdot x}\left[-2ip\cdot \partial+\partial^2\right]
\frac{1}{p^2+m^2}
\prod_{j=k+1}^N\left[V(x)\frac{1}{p^2+m^2}\right]
\pkt\ee
We now consider the sum over $N$; we split $N=k+l$ and
$(-1)^N=(-1)^k(-1)^l$.
The sum over $l$ is independent of $k$ and runs from $0$ to $\infty$
and, putting in the factor $(-1)^l$ we obtain
\bea \nonumber
&&\int d^4x e^{iQ_k\cdot x}\left[-2ip\cdot \partial+\partial^2\right]
\frac{1}{p^2+m^2}
\sum_{l=0}^\infty \prod_{j=1}^l\left[-V(x)\frac{1}{p^2+m^2}\right]
\\
&=&\int d^4x e^{iQ_k\cdot x}\left[-2ip\cdot \partial+\partial^2\right]
\frac{1}{p^2+m^2+V(x)}
\pkt\eea
Note that the sum starts with $l=0$, which corresponds to the case
$k=N$; in this case the product over $j$ reduces to $1$.
Now we do the analogous operations on the part to the
left of the insertion, using in the exponent $Q_k=q_1+\dots + q_k$;
we now can carry out the summation over k and we find finally
for the case that we
have taken into account the {\em first order expansion} of
{\em} one of
the denominators $(p+Q_k)^2+m^2$
\be
\int d^4x \frac{1}{p^2+m^2+V(x)}
\left[-2ip\cdot \partial+\partial^2\right] \frac{1}{p^2+m^2+V(x)}\pkt
\ee
Obviously part $-i2p\cdot \partial$ vanishes upon symmetric
integration over $p$. It also can be written as a boundary term
for the $x$ integration. If we want to obtain the second order
gradient term we have to take into account the
$Q_k^2$ term of the first order expansion, i.e.
\be
\int d^4x \frac{1}{p^2+m^2+V(x)}
\partial^2 \frac{1}{p^2+m^2+V(x)} \kma
\ee
the terms $-2ip\cdot \partial$ arising if two denominators
are expanded to first order, yielding
\be
\int d^4x \frac{1}{p^2+m^2+V(x)} (-2ip\cdot\partial)
\frac{1}{p^2+m^2+V(x)} (-2ip\cdot\partial)
\frac{1}{p^2+m^2+V(x)} \pkt
\ee
Here is included the term arising from expanding {\em one}
propagator to {\em second order}. Indeed this yields
\be
\frac{1}{p^2+m^2}\left[-2p\cdot Q_k-Q_k^2\right]\frac{1}{p^2+m^2}
\left[-2p\cdot Q_k-Q_k^2\right]\frac{1}{p^2+m^2} \kma
\ee
a term that is needed for obtaining the complete propagator
$1/(p^2+m^2+V(x))$ between the two insertions.
We now have the two-gradient term
\bea
\calg^{(2)}
&=&\frac{1}{2}\int\frac{d^4p}{(2\pi)^4}
\int d^4x \left\{
\frac{1}{p^2+m^2+V(x)}
\partial^2
\frac{1}{p^2+m^2+V(x)} \right.
\\ \nonumber
&& \left.+
\frac{1}{p^2+m^2+V(x)}
(-2ip\cdot\partial)
\frac{1}{p^2+m^2+V(x)} (-2ip\cdot\partial)
\frac{1}{p^2+m^2+V(x)}\right\} \pkt
\eea
The first term can be written, after one integration by parts
as
\be
\frac{1}{2}\int\frac{d^4p}{(2\pi)^4} \int d^4x
\frac{-1}{(p^2+m^2+V(x))^4}\left[\partial V(x)\right]^2 \pkt
\ee
In the second term we remark that the derivatives in the
first insertion act on the complete part to the right of it.
Therefore an integration by parts lets it act onto the
part to the left of it.
Using symmetric integration over $p$ the second part yields
\be
\frac{1}{2}\int\frac{d^4p}{(2\pi)^4} \int d^4x
\frac{p^2}{(p^2+m^2+V(x))^5}\left[\partial V(x)\right]^2 \pkt
\ee
Now we integrate with respect to $m^2$ to obtain the two-gradient
contribution to the one-loop effective action
\bea
\nonumber
S^{\rm eff}_{{\rm grad}, 2} &=&
\frac{1}{2}\int\frac{d^4p}{(2\pi)^4}
\int d^4x\left[\frac{1}{3}\frac{1}{(p^2+m^2+V(x))^3}
-\frac{1}{4}\frac{p^2}{(p^2+m^2+V(x))^3}\right]
\left[\partial V(x)\right]^2 \\
&=& \frac{1}{32\pi^2}
\int d^4x \frac{1}{m^2+V(x)} \frac{1}{12}
\left[\partial V(x)\right]^2 \kma
\eea
which coincides with the previous result Eq.~(\ref{grad2}).

The terms of the gradient expansion can be evaluated in a straightforward way.
We note, however, that the term $m^2+V(x)$ vanishes, depending on value 
of $\alpha$, at one or two points,
and that therefore the expressions are ill-defined a priori.
This is a reflection of the fact that the effective action has an 
imaginary part, due to the negative mode. An expansion of 
the effective action has to reflect this
feature. With an $m^2-i\epsilon$ prescription this becomes apparent.
When computing these terms we have used the principal value prescription for
$S^{\rm eff}_{{\rm grad}, 2} $ and taken the absolute value in the logarithm
appearing in $S^{eff}_{{\rm grad}, 0}$.


\newpage


\newpage

\begin{figure}
 \hbox to\hsize{
 \epsfig{file=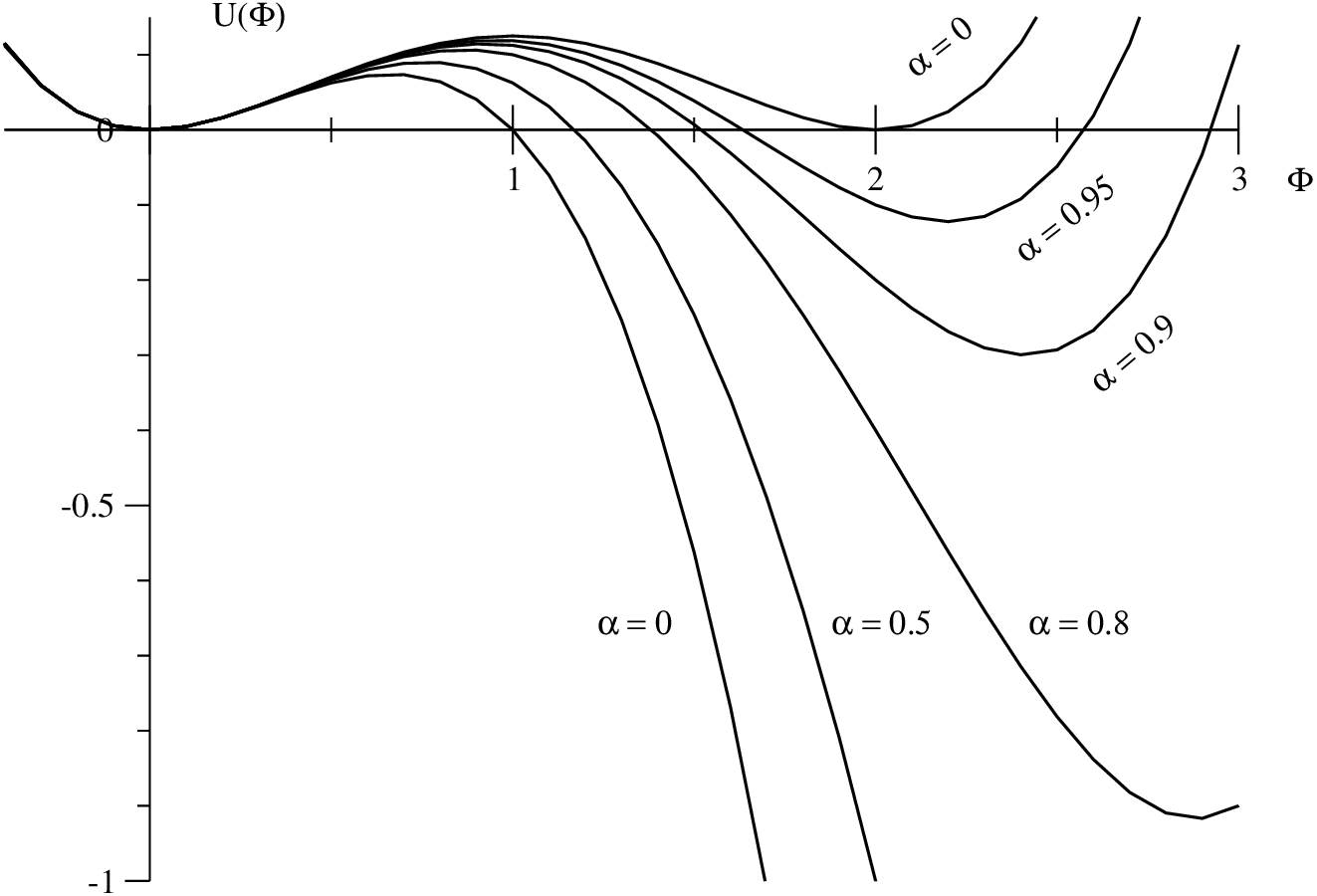,width=0.7\hsize,
 bbllx=0.7cm,bblly=1.5cm,bburx=20.5cm,bbury=19.5cm}\hss }
\labelcaption{potU}
{Potential $U(\Phi)$ in dimensionless form Eq.~(\ref{pot}).
The curves are labelled with the value of $\alpha $.}
\end{figure}

\begin{figure}
 \hbox to\hsize{
 \epsfig{file=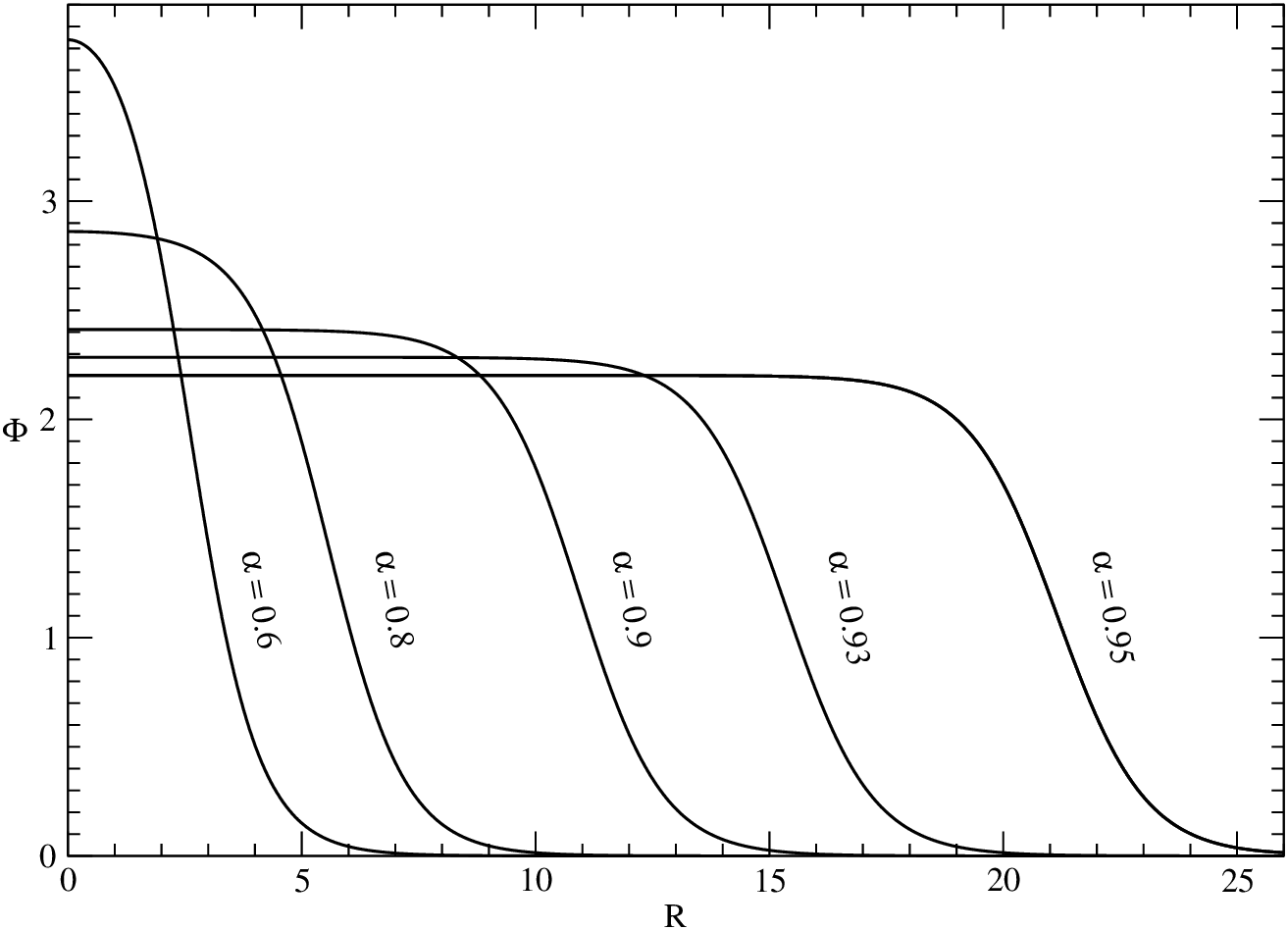,width=0.7\hsize,
 bbllx=0.7cm,bblly=1.5cm,bburx=20.5cm,bbury=19.5cm}\hss }
\labelcaption{phi}
{Bounce profiles for different $\alpha $.}
\end{figure}

\begin{figure}
\hbox to\hsize{
  \epsfig{file=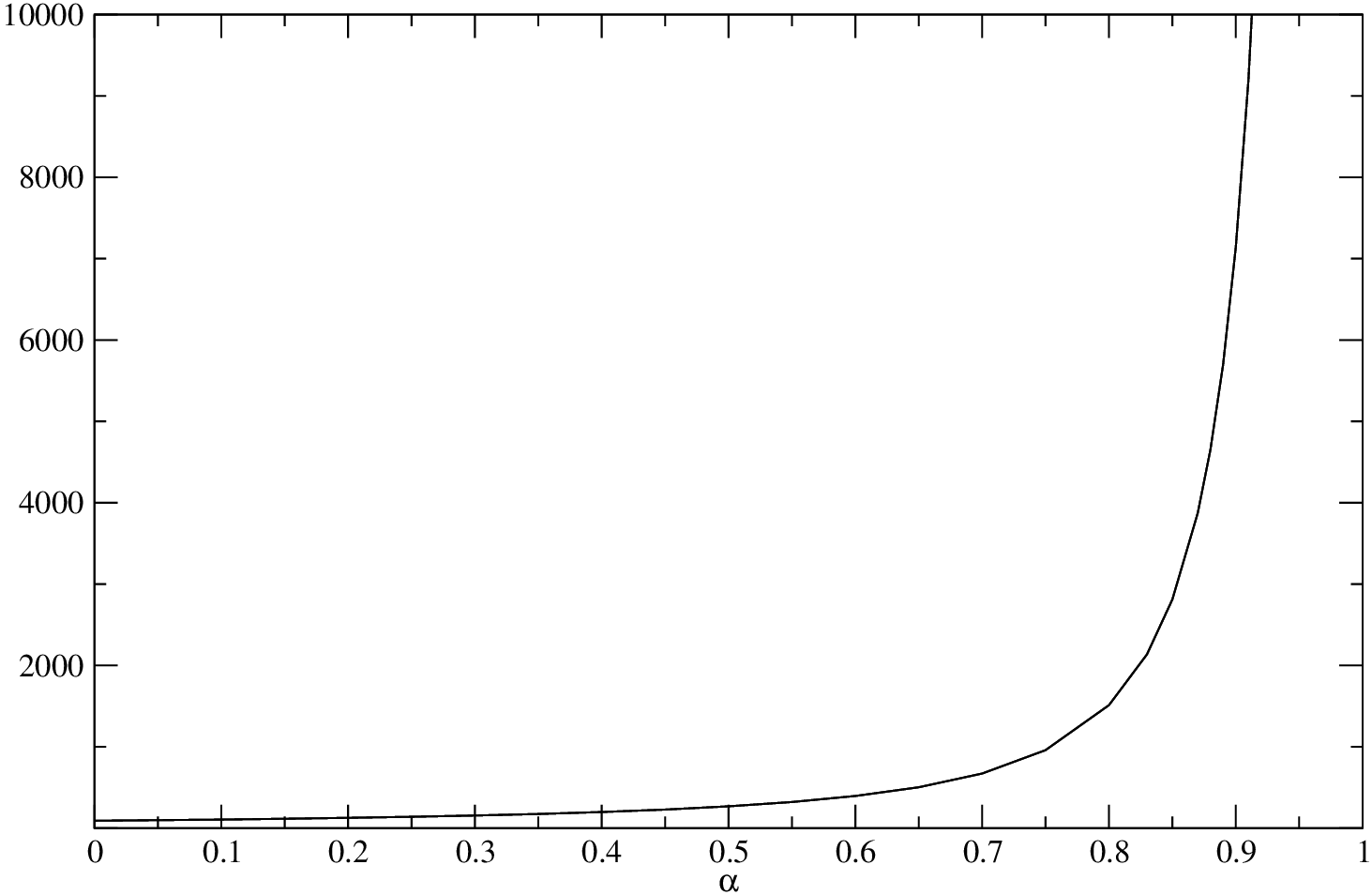,width=0.35\hsize,%
      bbllx=2.3cm,bblly=1.5cm,bburx=20.0cm,bbury=20.0cm}\hss
  \epsfig{file=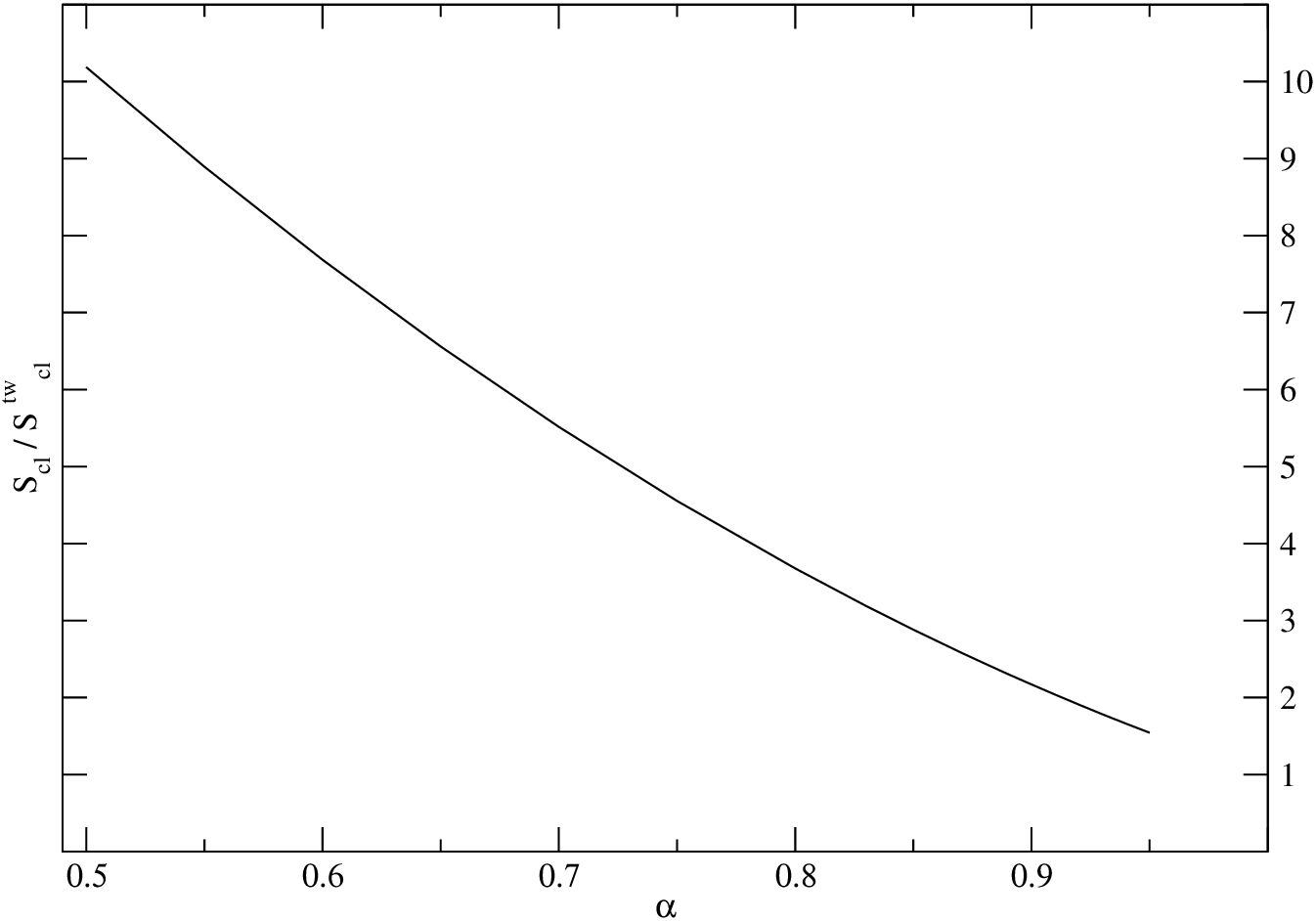,width=0.35\hsize,%
      bbllx=1.8cm,bblly=1.5cm,bburx=20.0cm,bbury=20.0cm}
  }
\labelcaption{sclass}{
Classical action $\tilde{S}_{cl}$ versus $\alpha$
(left) and the ratio $\tilde{S}_{cl}/\tilde{S}^{tw}_{cl}$
for $\alpha > 0.5$ (right).}
\end{figure}

\begin{figure}
\hbox to\hsize{
\epsfig{file=bounce.fig4.eps,width=0.7\hsize,
bbllx=0.7cm,bblly=1.5cm,bburx=20.5cm,bbury=19.5cm}\hss }
\labelcaption{Gplot}
{The ratio $G(\alpha,\beta)=S^{eff}_{1-loop}/ S_{cl}$ for $\beta=1$.}
\end{figure}

\begin{figure}
\hbox to\hsize{
\psfig{file=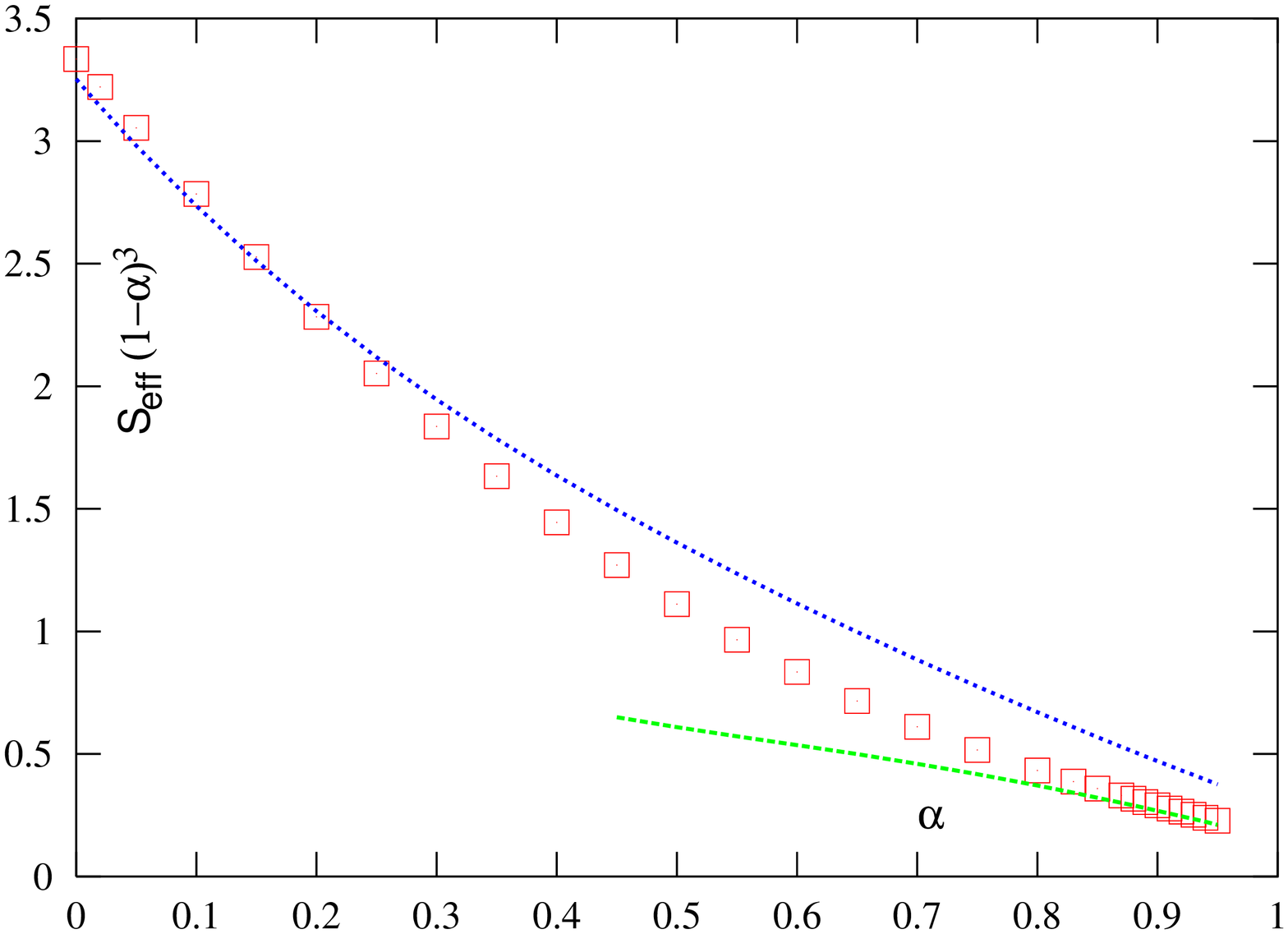,angle=-90,width=0.7\hsize,
bbllx=0.7cm,bblly=1.5cm,bburx=20.5cm,bbury=19.5cm}\hss }
\labelcaption{srg_sef}
{Our results for the effective action
$S^{eff}_{\rm 1-loop}$ (squares)
together with the perturbative part  $S^{eff}_{\rm 1-loop, p}$
(dotted line) and the leading parts of the gradient expansion
$S^{eff}_{{\rm grad}, 0+2}$ (dashed line, $\alpha=0.45-0.95$).
All shown quantities are multiplied by the factor $(1-\alpha)^3$.}
\end{figure}

\begin{figure}
\hbox to\hsize{
\epsfig{file=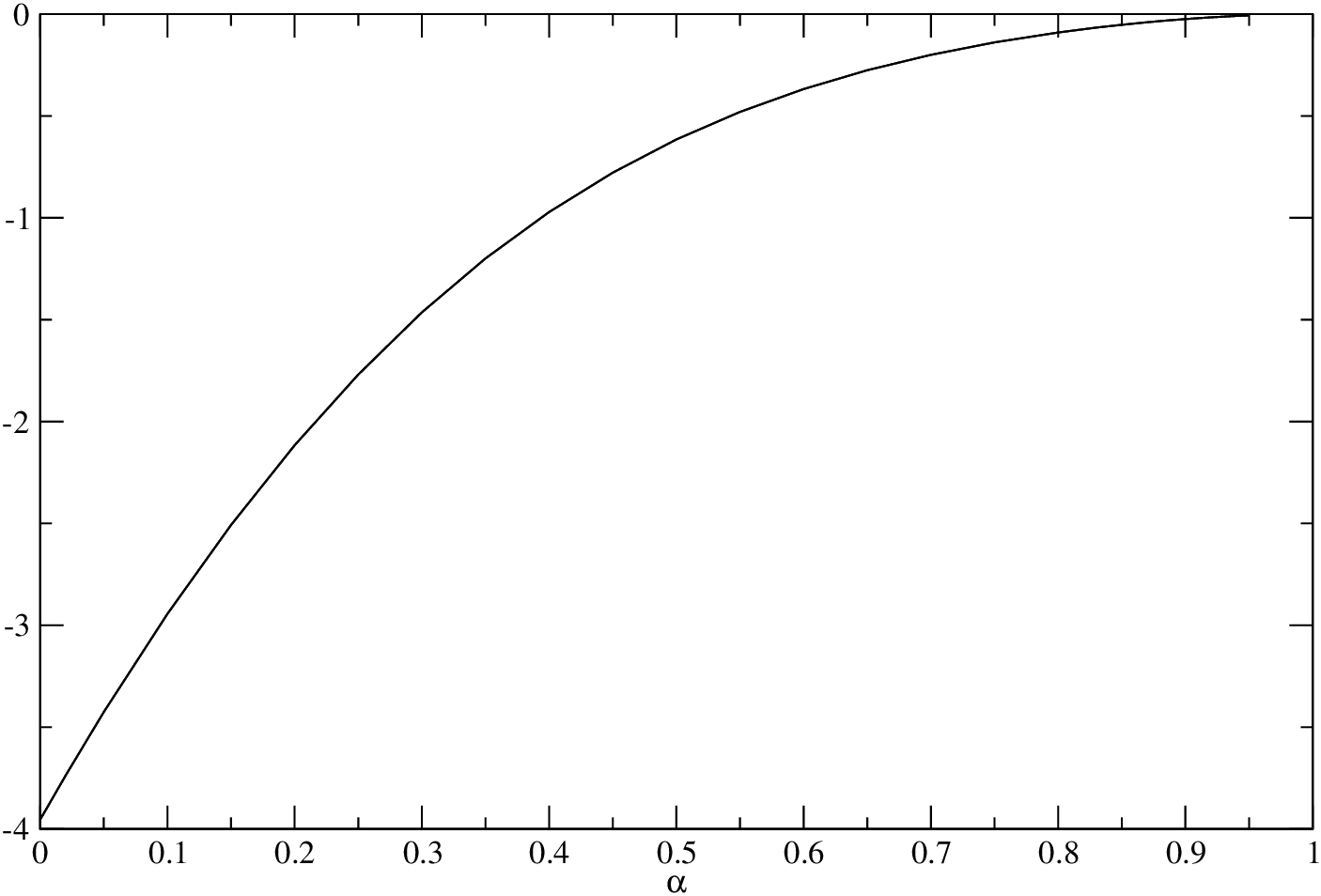,width=0.7\hsize,
bbllx=0.7cm,bblly=1.5cm,bburx=20.5cm,bbury=19.5cm}\hss }
\labelcaption{Omega}
{The negative mode energy as a function of $\alpha$}
\end{figure}

${}$
\begin{center}
\begin{tabular} {|c|l|l|l|l|}  \hline
$\alpha$&$S^{eff}_{\rm 1-loop,p}$& $ S^{eff}_{\rm 1-loop, n.p.}$& $S^{eff}_{1-loop}$
&$\tilde{S}_{cl}$\\  \hline
0.00&3.253E+00&  8.216E-02&3.335E+00&9.086E+01\\
0.02&3.337E+00&  8.498E-02&3.422E+00&9.355E+01\\
0.05&3.478E+00&  8.422E-02&3.562E+00&9.787E+01\\
0.10&3.752E+00&  6.737E-02&3.819E+00&1.059E+02\\
0.15&4.089E+00&  2.654E-02&4.115E+00&1.153E+02\\
0.20&4.504E+00& -4.501E-02&4.459E+00&1.263E+02\\
0.25&5.021E+00& -1.564E-01&4.865E+00&1.394E+02\\
0.30&5.672E+00& -3.201E-01&5.351E+00&1.552E+02\\
0.35&6.499E+00& -5.539E-01&5.946E+00&1.744E+02\\
0.40&7.571E+00& -8.836E-01&6.687E+00&1.983E+02\\
0.45&8.984E+00& -1.348E+00&7.637E+00&2.286E+02\\
0.50&1.089E+01& -2.006E+00&8.889E+00&2.681E+02\\
0.55&1.356E+01& -2.958E+00&1.060E+01&3.211E+02\\
0.60&1.741E+01& -4.371E+00&1.303E+01&3.951E+02\\
0.65&2.326E+01& -6.560E+00&1.670E+01&5.033E+02\\
0.70&3.277E+01& -1.015E+01&2.261E+01&6.720E+02\\
0.75&4.966E+01& -1.659E+01&3.306E+01&9.589E+02\\
0.80&8.382E+01& -2.969E+01&5.413E+01&1.512E+03\\
0.83&1.240E+02& -4.512E+01&7.887E+01&2.136E+03\\
0.85&1.686E+02& -6.233E+01&1.062E+02&2.809E+03\\
0.87&2.409E+02& -9.038E+01&1.506E+02&3.874E+03\\
0.88&2.950E+02& -1.114E+02&1.836E+02&4.655E+03\\
0.89&3.684E+02& -1.401E+02&2.283E+02&5.699E+03\\
0.90&4.711E+02& -1.803E+02&2.907E+02&7.140E+03\\
0.91&6.199E+02& -2.390E+02&3.809E+02&9.198E+03\\
0.92&8.455E+02& -3.284E+02&5.171E+02&1.227E+04\\
0.93&1.207E+03& -4.724E+02&7.347E+02&1.711E+04\\
0.94&1.829E+03& -7.209E+02&1.109E+03&2.531E+04\\
0.95&3.008E+03& -1.188E+03&1.820E+03&4.061E+04\\
\hline
\end{tabular}
\vglue 0.4cm
Tab.~1. Numerical results for classical action
and one loop effective action.
\end{center}

\newpage
\begin{center}
\begin{tabular} {|c|l|l|l|}  \hline
$\alpha$&$A^{(1)}_{fin}$&$A^{(2)}_{fin}$&$a^{(2)}$\\  \hline
0.00&5.178E+00& -2.654E+00&1.036E+01 \\
0.02&5.379E+00& -2.591E+00&1.055E+01 \\
0.05&5.706E+00& -2.499E+00&1.085E+01 \\
0.10&6.325E+00& -2.358E+00&1.144E+01 \\
0.15&7.060E+00& -2.235E+00&1.215E+01 \\
0.20&7.942E+00& -2.133E+00&1.300E+01 \\
0.25&9.013E+00& -2.059E+00&1.405E+01 \\
0.30&1.033E+01& -2.020E+00&1.536E+01 \\
0.35&1.199E+01& -2.024E+00&1.702E+01 \\
0.40&1.410E+01& -2.085E+00&1.916E+01 \\
0.45&1.686E+01& -2.219E+00&2.199E+01 \\
0.50&2.056E+01& -2.453E+00&2.585E+01 \\
0.55&2.570E+01& -2.825E+00&3.127E+01 \\
0.60&3.311E+01& -3.399E+00&3.921E+01 \\
0.65&4.438E+01& -4.286E+00&5.147E+01 \\
0.70&6.269E+01& -5.701E+00&7.175E+01 \\
0.75&9.526E+01& -8.109E+00&1.085E+02 \\
0.80&1.613E+02& -1.269E+01&1.848E+02 \\
0.83&2.391E+02& -1.778E+01&2.761E+02 \\
0.85&3.256E+02& -2.321E+01&3.790E+02 \\
0.87&4.660E+02& -3.170E+01&5.479E+02 \\
0.88&5.711E+02& -3.788E+01&6.753E+02 \\
0.89&7.137E+02& -4.609E+01&8.493E+02 \\
0.90&9.134E+02& -5.735E+01&1.094E+03 \\
0.91&1.203E+03& -7.333E+01&1.453E+03 \\
0.92&1.643E+03& -9.699E+01&1.999E+03 \\
0.93&2.347E+03& -1.341E+02&2.883E+03 \\
0.94&3.561E+03& -1.963E+02&4.417E+03 \\
0.95&5.861E+03& -3.118E+02&7.359E+03 \\
\hline
\end{tabular}
\vglue 0.4cm
Tab.~2. Numerical results for the first
and second order contribution coefficients.
\end{center}

\begin{thebibliography}{}

\bibitem{La67}
J. S. Langer,
Ann.Phys. (N.Y.) {\bf 41} (1967) 108.

\bibitem{La69}
J. S. Langer,
Ann.Phys. (N.Y.) {\bf 54} (1969) 258.

\bibitem{vko75}
M.~B.~Voloshin, I.~Y.~Kobzarev and L.~B.~Okun,
Sov.\ J.\ Nucl.\ Phys.\  {\bf 20} (1975) 644.

\bibitem{Co77}
S. Coleman,
Phys. Rev. {\bf D15} (1977) 2929.

\bibitem{CaCo77}
C.~G.~Callan and S.~R.~Coleman,
Phys.\ Rev.\ D {\bf 16} (1977) 1762.

\bibitem{Co79}
S. Coleman,
'The Uses of Instantons' in "The Whys of Subnuclear Physics",
A. Zichichi ed., Plenum Press, New York 1979.

\bibitem{ks84}
V. G. Kiselev and K. G. Selivanov,
Sov. Phys. JTEP Lett. {\bf 39} (1984) 85.

\bibitem{ks86}
V. G. Kiselev and K. G. Selivanov,
Sov. J. Nucl. Phys. {\bf 43} (1986) 153.

\bibitem{se88}
K. G. Selivanov,
Sov. Phys. JETP {\bf 67} (1988) 1548.

\bibitem{baki93}
J.~Baacke and V.~G.~Kiselev,
Phys.\ Rev.\ D {\bf 48}(1993) 5648,
[arXiv: hep-ph/9308273].

\bibitem{Carson:1990jm}
L.~Carson, X.~Li, L.~D.~McLerran and R.~T.~Wang,
Phys.\ Rev.\ D {\bf 42} (1990) 2127.

\bibitem{baju93}
J.~Baacke and S.~Junker,
Phys.\ Rev.\ D {\bf 49} (1994) 2055,
[arXiv:hep-ph/9308310]; \\
Phys.\ Rev.\ D {\bf 50} (1994) 4227,
[arXiv:hep-th/9402078].

\bibitem{Baacke:1995bw}
J.~Baacke,
Phys.\ Rev.\ D {\bf 52} (1995) 6760,
[arXiv:hep-ph/9503350].

\bibitem{ba91}
J. Baacke, Acta Phys. Pol. {\bf B22} (1991) 127
and references therein.

\bibitem{ba92}
J. Baacke,
Z. Phys. {\bf C53} (1992) 402.

\bibitem{Baacke:1994bk}
J.~Baacke and T.~Daiber,
Phys.\ Rev.\ D {\bf 51} (1995) 795,
[arXiv:hep-th/9408010].

\bibitem{Diakonov:1995xz}
D.~Diakonov, M.~V.~Polyakov, P.~Sieber, J.~Schaldach and K.~Goeke,
Phys.\ Rev.\ D {\bf 53} (1996) 3366,
[arXiv:hep-ph/9502245].

\bibitem{Graham:2002xq}
N.~Graham, R.~L.~Jaffe, V.~Khemani, M.~Quandt, M.~Scandurra and H.~Weigel,
arXiv: hep-th/0207120.

\bibitem{Bordag:2002sa}
M.~Bordag,
Phys.\ Rev.\ D {\bf 67} (2003) 065001,
[arXiv:hep-th/0211080].

\bibitem{Hellmund:1993uh}
M.~Hellmund, J.~Kripfganz and M.~G.~Schmidt,
Phys.\ Rev.\ D {\bf 50} (1994) 7650,
[arXiv:hep-ph/9307284].

\bibitem{Fliegner:1993wh}
D.~Fliegner, M.~G.~Schmidt and C.~Schubert,
Z.\ Phys.\ C {\bf 64} (1994) 111,
[arXiv:hep-ph/9401221].

\bibitem{Fliegner:1997rk}
D.~Fliegner, P.~Haberl, M.~G.~Schmidt and C.~Schubert,
Annals Phys.\  {\bf 264} (1998) 51,
[arXiv:hep-th/9707189].

\bibitem{mr99}
G.~Munster and S.~Rotsch,
Eur.\ Phys.\ J.\ C {\bf 12} (2000) 161,
[arXiv:cond-mat/9908246].

\bibitem{Caro:1993fs}
J.~Caro and L.~L.~Salcedo,
Phys.\ Lett.\ B {\bf 309} (1993) 359.

\bibitem{rubakov84}
V.~A.~Rubakov,
Nucl.\ Phys.\ B {\bf 245} (1984) 481.

\bibitem{surig97}
A.~Surig,
Phys.\ Rev.\ D {\bf 57} (1998) 5049,
[arXiv:hep-ph/9706259].

\bibitem{levkov02}
D.~Levkov, C.~Rebbi and V.~A.~Rubakov,
Phys.\ Rev.\ D {\bf 66} (2002) 083516,
[arXiv:gr-qc/0206028].

\bibitem{CGM78}
S. Coleman, V. Glaser and A. Martin,
Commun. Math. Phys. {\bf 58} (1978) 211.

\bibitem{cw84}
P.~Candelas and S.~Weinberg,
Nucl.\ Phys.\ B {\bf 237} (1984) 397.

\bibitem{DLHLL}
M. Dine, R. G. Leigh, P. Huet, A. Linde, D. Linde,
Phys.Rev. {\bf D46} (1992) 550;
Phys. Lett. {\bf B283} (1992) 319.

\bibitem{theorem}
R. F. Dashen, B. Hasslacher, and A. Neveu,
Phys.Rev. {\bf D10} (1974) 4114;
I. M. Gel'fand and A. M. Yaglom, J. Math. Phys. {\bf 1} (1960) 48;
R. H. Cameron and W. T. Martin, Bull. Am. Math. Soc. {\bf51 } (1945) 73;
J. H. van Vleck, Proc. Nat. Acad. Sci. {\bf 14} (1928) 178.

\bibitem{zurmuel}
See e.g.:
R.~Zurm\"uhl, Praktische Mathematik f\"ur Ingenieure und Physiker,
Springer, 1984.

\bibitem{psbook}
M.~E.~Peskin, D.~V.~Schroeder, Introduction to quantum field theory,
Perseus Books, 1995.

\bibitem{kr85}
R.~V.~Konoplich and S.~G.~Rubin,
Sov. J. Nucl. Phys. {\bf 42} (1985) 810.

\bibitem{stw99}
A.~Strumia, N.~Tetradis and C.~Wetterich,
Phys.\ Lett.\ B {\bf 467} (1999) 279,
[arXiv:hep-ph/9808263].

\bibitem{mst00}
G.~Munster, A.~Strumia and N.~Tetradis,
Phys.\ Lett.\ A {\bf 271} (2000) 80,
[arXiv: cond-mat/0002278].

\end{thebibliography}
\end{document}